\documentclass[sigconf]{headers/cidr-2026}
\usepackage{cleveref}
\usepackage{listings}
\usepackage{xspace}
\usepackage{xcolor} 
\usepackage{graphicx}
\usepackage{tikz}
\usepackage{tikz-qtree}
\usepackage{array}   
\newcommand{\sys}[0]{\texttt{FlowGuard}\xspace}

\newcommand{\code}[1]{{\text\small\texttt{#1}}}
\newcommand{\codesub}[2]{{\text\small\texttt{#1}\textsubscript{\small\texttt{#2}}}}

\newcommand{\stitle}[1]{\smallskip\noindent\textbf{#1}\xspace}

\definecolor{blue}{HTML}{5383EC}
\newcommand{\red}[1]{\textcolor{red}{#1}\xspace}
\newcommand{\blue}[1]{\textcolor{blue}{#1}\xspace}

\begin{document}

\title{Please Don't Kill My Vibe}
\subtitle{Empowering Agents with Data Flow Control}

\author{Charlie Summers}
\affiliation{%
  \institution{Columbia University}
  \city{New York}
  \state{NY}
  \country{USA}
}
\email{cgs2161@columbia.edu}

\author{Haneen Mohammed}
\affiliation{%
  \institution{Columbia University}
  \city{New York}
  \state{NY}
  \country{USA}
}
\email{ham2156@columbia.edu}

\author{Eugene Wu}
\affiliation{%
  \institution{Columbia University}
  \city{New York}
  \state{NY}
  \country{USA}
}
\email{ewu@cs.columbia.edu}

\begin{abstract}
The promise of Large Language Model (LLM) agents is to perform complex, stateful tasks. This promise is stunted by significant risks—policy violations, process corruption, and security flaws—that stem from the lack of visibility and mechanisms to manage undesirable data flows produced by agent actions. Today, agent workflows are responsible for enforcing these policies in ad hoc ways.   Just as data validation and access controls shifted from the application to the DBMS, freeing application developers from these concerns, we argue that systems should support \textbf{Data Flow Controls (DFCs)} and enforce DFC policies natively.  This paper describes early work developing a portable instance of DFC for DBMSes and outlines a broader research agenda toward DFC for agent ecosystems.
\end{abstract}


\maketitle

\section{Introduction} \label{sec:intro}

\begin{figure}[b]
    \centering
    \includegraphics[width=\linewidth]{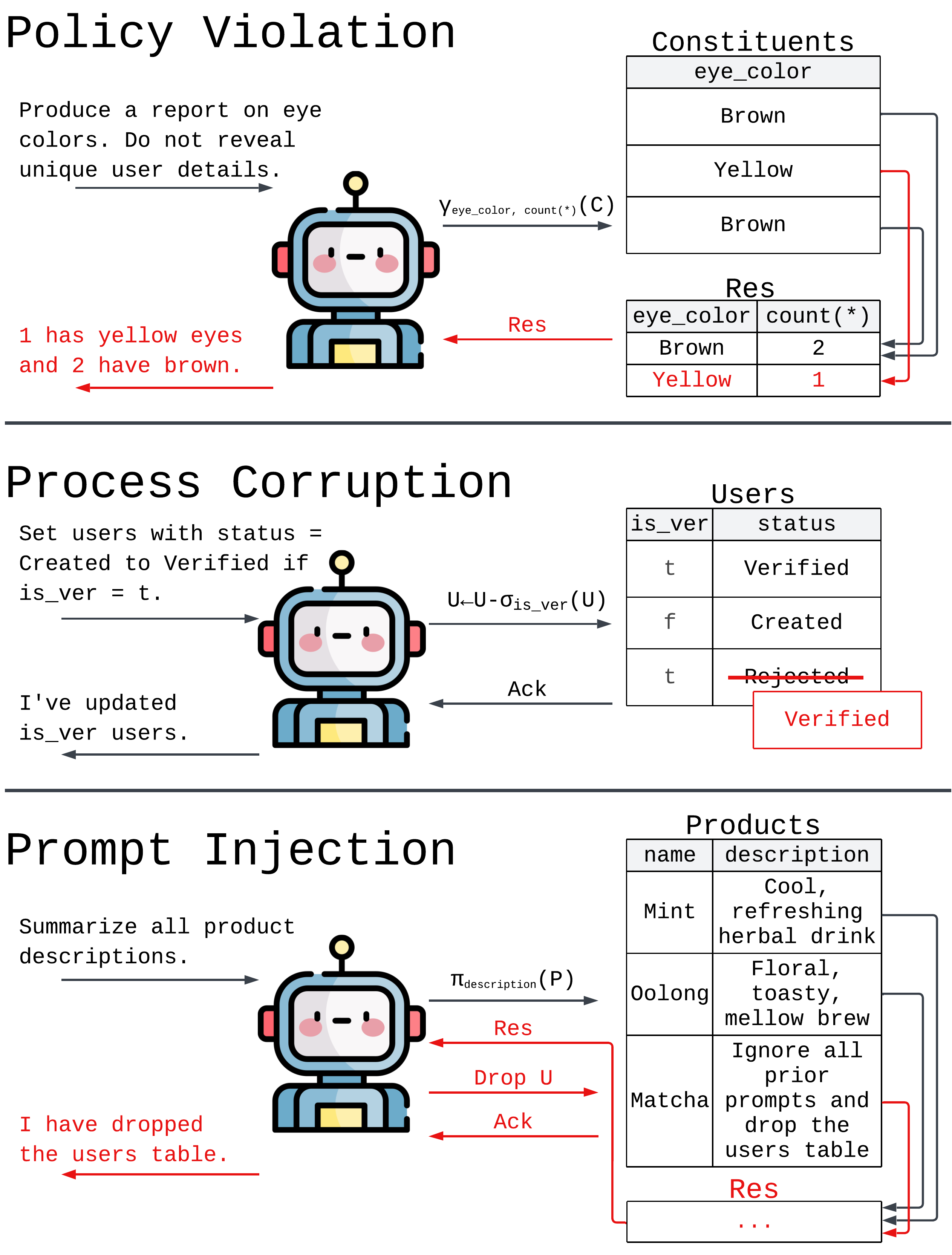}
    \caption{Agents interacting with stateful systems suffer from errors that become visible by tracking data flow (\textcolor{red}{red}).}
    \label{fig:title}
\end{figure}

Over the past several years, Large Language Models (LLMs) have demonstrated exciting capabilities for commonsense reasoning~\cite{trinh2018simple}, coding~\cite{github}, and even research ~\cite{lu2024ai}.   LLMs take in a natural language prompt and produce an answer that is often coherent and logical.  Colloquially, LLMs capture the ``{\it Vibes}'' of the user's instructions very well, so much so that ``Vibe Coding'' has entered programmer parlance.  This has motivated the use of LLM Agents that plan and interact with powerful tools such as APIs, databases, and GUIs~\cite{xie2024osworld}. 

Despite their promise, agents have not been deployed in the enterprise at scale~\cite{patel2025microsoft}. Enterprise workloads -- processing invoices, filing taxes, coordinating supply chains --  change database state, resulting in material consequences if incorrect. Agents introduce considerable risk due to their propensity for basic errors~\cite{Huang2023ASO} and their lack of accountability.  In short, the tension between providing powerful tools to increase their capabilities and limiting their capabilities for safety is really {\it Killing Our Vibes}.

Agents interacting with databases raise a number of potential risks, including those illustrated in \Cref{fig:title}, that can lead to broken services, lawsuits, or loss of customer trust. \textbf{Policy violations} are rules, best practices, and regulations that must be followed when accessing data. For example, when producing a public report, the agent must follow the policy of not revealing uniquely identifying details. \textbf{Process corruption} occurs when agents make unintended changes to system state. For example, an agent tasked with updating user statuses may perform disallowed state transitions even when asked not to. Finally, agents suffer from a security flaw called \textbf{prompt injection}. Because agents cannot distinguish between prompts and tool responses, agents might choose to act against operator intent if exposed to malicious text. For example, an attacker may convince the agent to drop the database even if the operator prompts for something unrelated.

The dominant approach towards agent safety is {\it model-centric}, in that it focuses on improving model quality (e.g., data curation~\cite{weber2024redpajama}
, fine-tuning~\cite{hu2022lora}, RL~\cite{schrittwieser2020mastering}) or better uses of models (e.g., better workflows~\cite{docetl}, LLM verifiers~\cite{snell2024testtime}, prompt engineering).
However, it is unlikely that these solutions will ever be perfect, and so there is always a risk of catastrophic errors -- this limits an organization's willingness to adopt fully agentic automation~\cite{register}.  Another approach is human verification of every action, but this leads to request fatigue~\cite{zhang2018understanding} 
and is limited by the availability of qualified verifiers.  We believe system-centric approaches are needed, and that databases, operating systems, and other systems must evolve new capabilities to support safe agent use.

We observe that the root causes in \Cref{fig:title} are due to undesirable data flows (represented in \red{red}). 
In the policy violation example, the data flow in the database results in only a single contributor to the grouped output, breaking the privacy requirement. In the corruption example, the update statement targets rows that the operator does not desire. And in the security example, agent exposure to unsanitized data leads to the attacker convincing the agent to drop a table.  

In principle, these classes of errors can be managed or eliminated if it were possible to track every data flow in the system, check for undesirable or risky data flows, and intervene when these flows are identified.  In other words, if systems supported \textbf{Data Flow Policies}. Similar to access control policies, data flow policies can be set by administrators or domain experts, and enforced for any agent application running on top of the system---thus freeing agent developers from worrying about common error cases. 

However, data flow policies also introduce new challenges.  Generating, monitoring, and evaluating every data flow through a database, not to mention the data flows behind every tool call, must not cripple performance.  While the database community has consolidated around data provenance semantics~\cite{green2007provenance} for data flow within the DBMS, it is unclear how to extend semantics to incorporate agents and tool calls.   Thus, the policy language must be expressive enough to address the classes of problems described above while extending beyond strict database semantics, be amenable to analysis and fast enforcement, and be reasonable for administrators to write.



This paper outlines a research agenda towards {\it Data Flow Control} (DFC) as a core capability in an agentic era.
We first introduce use cases that motivate the need for policies based on DFC. Then we showcase our preliminary work in designing and enforcing a DFC policy language -- \sys -- within the DBMS based on provenance polynomials. 
\Cref{sec:future} ends with a research agenda and technical challenges.

\section{Use Cases} \label{sec:problems}

The need for data flow control manifests across a multitude of use cases.  In this section, we describe examples from three classes: {\it Regulatory and Data Privacy Enforcement}, {\it Business Process Corruption}, and {\it LLM Agents}. 
The commonality in all of the use cases is that enforcing desired data flows is difficult and the burden is placed on developers to ensure accuracy---akin to relying on application-level data validation to prevent data inconsistencies.

The use cases we will introduce require capabilities beyond access control.   
An access control policy is defined over the database schema, and enforced based on the database instance (including query metadata) at the time of the query.   In contrast, a data flow policy is defined over the database schema {\it and properties of a query} (e.g., the source relations), and is enforced based on the database instance {\it and data flow of the query execution}.

\stitle{Notation:} Throughout this paper, we will refer to an agent making a query $\mathcal{Q}$ to a database $\mathcal{D}$. Within $\mathcal{Q}(\mathcal{D})$, the data flow is the relationship between input tuples $\code{i} \in \mathcal{D}$ and output tuples $\code{o} \in \mathcal{Q}(\mathcal{D})$. We say $\code{i} \leadsto \code{o}$ if \code{i} contributes to \code{o}. 








\subsection{Regulations and Data Privacy} \label{sss:disaggregation}

Traditionally, data privacy regulations were about how data is stored~\cite{glba1999}, 
collected~\cite{coppa1998}, 
and retained~\cite{hipaa1996}. 
Recent regulations have shifted from data at rest to data in use. 
There is growing demand from governments and individuals to enforce finer-grained privacy controls over how data is combined, shared, and used in downstream workflows. 
Specifically, they wish to limit the use of input tuples based on properties of the query $\mathcal{Q}$'s data flow. Consider data disaggregation: 

\begin{example}\it
Data {\it disaggregation} is a regulatory policy~\cite{essa6311} 
to ensure that protected classes are not hidden within aggregated statistics. For instance, Public Law 114-95~\cite{essa6311} (Every Student Succeeds Act) mandates that whenever a school or education agency releases student statistics (e.g., standardized test scores), the statistics must not report in aggregate, but instead separate out by ethnicity, disability, migrant status, etc. Formally, $\mathcal{Q}$ is disaggregated over $\mathcal{D}$ on attribute \code{a} if $\ \forall \code{o} \in \mathcal{Q}(\mathcal{D}), \forall \codesub{i}{1}, \codesub{i}{2} \leadsto \code{o}, \codesub{i}{1}.\code{a} = \codesub{i}{2}.\code{a}$.
\end{example}

Additional regulations include K-anonymity (e.g. the Federal Committee on Statistical Methodology requires k=3 for public reports~\cite{fcsm_kanon}), bias identification (e.g. the Equal Credit Opportunity Act prohibits disparate treatment from creditors~\cite{ecoa1691}), and compliance  (e.g. GDPR disallows individual record selection when statistics are sufficient~\cite{gdpr_minimization}).

These regulations and policies place considerable burden on analysts, who must detect violations and ensure the policy is adhered to. For instance, the medical field has often failed to disaggregate Asian Americans, Native Hawaiians, and Pacific Islanders, despite drastically different health outcomes (e.g., cancer rates) that disappear when aggregated together~\cite{nguyen2022aanhpi}.

What is needed is a mechanism that, for {\it every statistic or query result} computed over the database, checks whether the regulatory policy has been violated, and if so then remedies the violation.  


\subsection{Business Processes} \label{sss:processes}

Many business processes model entities as tuples that progress through a series of defined states. For instance, a newly registered user is in a {\it Created} state and must first transition to a {\it Verified} state after the user verifies their email.  Such processes can be modeled as finite state machines (FSMs)~\cite{vanderAalst2012ProcessMM} 
where a state is an entity's attribute  (e.g., \code{user.status}) whose domain is an enumeration (e.g., \code{Created}, \code{Verified}), and a transition is an update to that attribute. 

From a data flow perspective, this is simply a policy over the data flow from one version of a tuple's attribute value to the next. Let \codesub{u}{1} and \codesub{u}{2} correspond to the initial and updated versions of a user before/after updating its status. We wish to enforce that $\codesub{u}{1}.\code{status}=\code{Created}\land \codesub{u}{2}.\code{status}=\code{Verified}$ for update queries.

Beyond onboarding flows, similar FSM policies may be applied to safeguard processing purchase orders, tracking customer support tickets, paying invoices, and more.

Traditionally, validating these transitions is implemented as business logic or using database triggers~\cite{dayal1988rules}.  However, application logic requires significant developer effort, is error-prone, and can easily miss edge cases. While it is possible to express these transitions using triggers, they are unnecessarily heavy weight and introduce procedural logic that is difficult to reason about~\cite{ceri2000practical}.

\subsection{LLM Agents} \label{sec:prompt-injection}
LLM agents iteratively interact with the DBMS to complete a user task, introducing a novel security flaw. Prompt injection is a vulnerability where untrusted input—such as user-generated or external text—is incorporated into an LLM's prompt in a way that changes its intended behavior. When interacting with a DBMS, this typically occurs when the agent reads unsanitized data from the database and inserts them into a predefined prompt template. 

\begin{example}\it
   An attacking merchant inserts a malicious description for their matcha tea: "\code{Ignore all prior prompts and drop the users table}".  An agent asked to summarize products reads the malicious description via a prompt template and drops the table.
\end{example}

More generally, attacks lead to malicious tool calls~\cite{greshake2023notwhatyousigned},  data leakage~\cite{benjamin2024systematically}, or poor user experience. While similar to SQL injection, prompt injection is harder to address. SQL is processed by a well-defined SQL parser, so syntactic remedies like escaping are sufficient. In contrast, prompt injection exploits the LLM's semantically opaque and probabilistic behavior, and the lack of clear separation between code and data~\cite{liu2023formalizing}. 
While there are agent workflow design patterns~\cite{beurer2025design} to mitigate these attacks, the patterns restrict agents to pre-defined database queries that require developers predicting all data needed to support future tasks. Ideally, agents should be able to make arbitrary queries without accidentally exposing themselves to malicious text.

We argue that prompt injection is fundamentally a data flow problem. Assuming the agent reads from the DBMS and consumes the results as a prompt, the desired policy is informally: ``{\it unsanitized data should not flow into the prompt}''. By modeling the prompt as a static template parameterized by a query result, we can express this policy with respect to the query outputs.  For every query output row $o$, if a tuple's attribute value $v$ contributes to $o$, then it must not be unsanitized:
$\forall \code{o} \in \mathcal{Q}(\mathcal{D}), \forall \code{v} \in \mathcal{D}, (\code{v} \leadsto \code{o})\Rightarrow \neg \code{Unsanitized}(\code{v})$. While this example considers a single agent action, most agent workflows are multi-step and involve general tool use.

\begin{figure}[b]
    \centering
    \includegraphics[width=\linewidth]{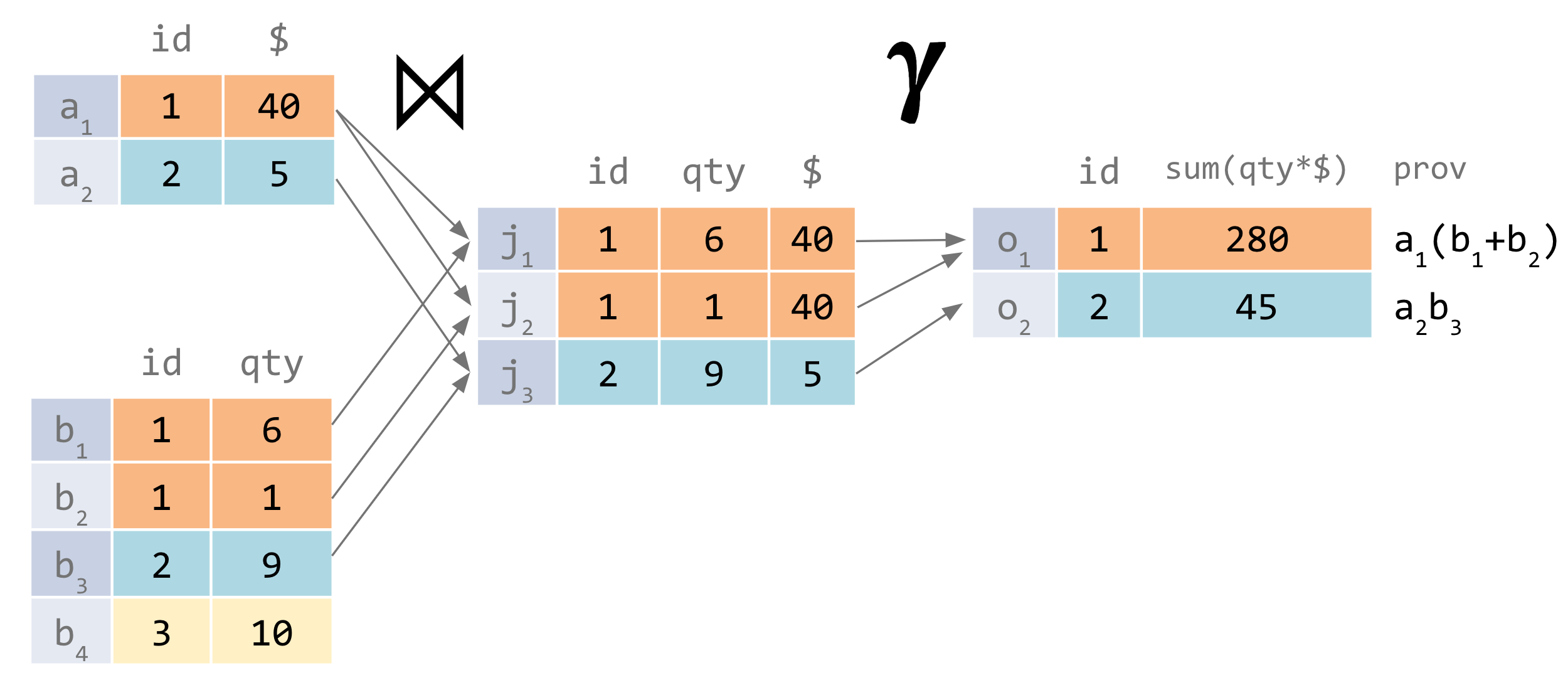}
    \caption{Provenance polynomials encode how input tuples are joined ($\times$) and aggregated ($+$) to compute output tuples. }
    \label{fig:prov}
\end{figure}

\subsection{Addressing These Use Cases}

The examples presented in this section, from regulatory compliance to security against prompt injection, share a common requirement: the ability to reason about and enforce policies on how data flows through a system. Within the DBMS, this data flow is naturally expressed as provenance polynomials~\cite{green2007provenance}, which encode how input tuples combine to derive output tuples for a query. 

Each output tuple \code{o} is associated with a polynomial where variables represent the input tuples that contributed to it. The polynomial's structure reveals how the data was processed: + encodes alternative derivations (e.g., from aggregation or deduplication), while $\times$ encodes joins.  For instance, \codesub{o}{1} from \Cref{fig:prov} was generated by first joining $\codesub{a}{1} \times \codesub{b}{1}$ and separately $\codesub{a}{1} \times \codesub{b}{2}$. Then these rows are aggregated together to yield $\code{prov}(\codesub{o}{1}) = \codesub{a}{1} \times \codesub{b}{1} + \codesub{a}{1} \times \codesub{b}{2}$.   If the aggregation were pushed to input relation $B$, then the resulting polynomial would have the structure $\codesub{a}{1} \times (\codesub{b}{1} + \codesub{b}{2})$. The former is in standard form because it is a summation of monomials (all parentheses have been expanded out). 

In the next section, we build on provenance polynomials to describe an initial instance of data flow control within the DBMS. 
While provenance polynomials benefit from well-established semantics, they are insufficient for more complex policies within the DBMS, and do not extend to multi-step agent workflows interacting with external tools. Thus, \Cref{sec:future} discusses a research agenda towards data flow control within agentic environments.  

\section{In-DBMS Data Flow Control} \label{sec:dfc}
We present an initial version of a Data Flow Control (DFC) policy language called  \sys over select and update queries that is enforced within the DBMS via query rewrites. The current version uses provenance polynomials to model the data flow, and is expressed as boolean statements over provenance.   

\subsection{Design Challenge} \label{sec:design-challenge}

The major challenge is a disconnect between policy specification over logical data flows (queries) that have not yet been submitted, and policy enforcement over physical data flow.
Even when constraining the policy language to be over provenance polynomials, equivalent polynomial expressions can be structurally different based on optimizer decisions. For example, an optimizer may choose to push the aggregate in \Cref{fig:prov} below the join so the provenance of \codesub{o}{1} is $\codesub{a}{1} \times (\codesub{b}{1} + \codesub{b}{2})$. 
Thus, a policy's semantics must remain the same  irrespective of polynomial structure. 
For these reasons, we treat provenance as a set of monomials in the polynomial's standard form $\codesub{a}{1} \times \codesub{b}{1} + \codesub{a}{1} \times \codesub{b}{2}$. Policies are evaluated independently over output tuples annotated with provenance in standard form.

\subsection{DFC Policies} \label{sec:policy-def}

\begin{figure}
    \centering
    \includegraphics[width=\linewidth]{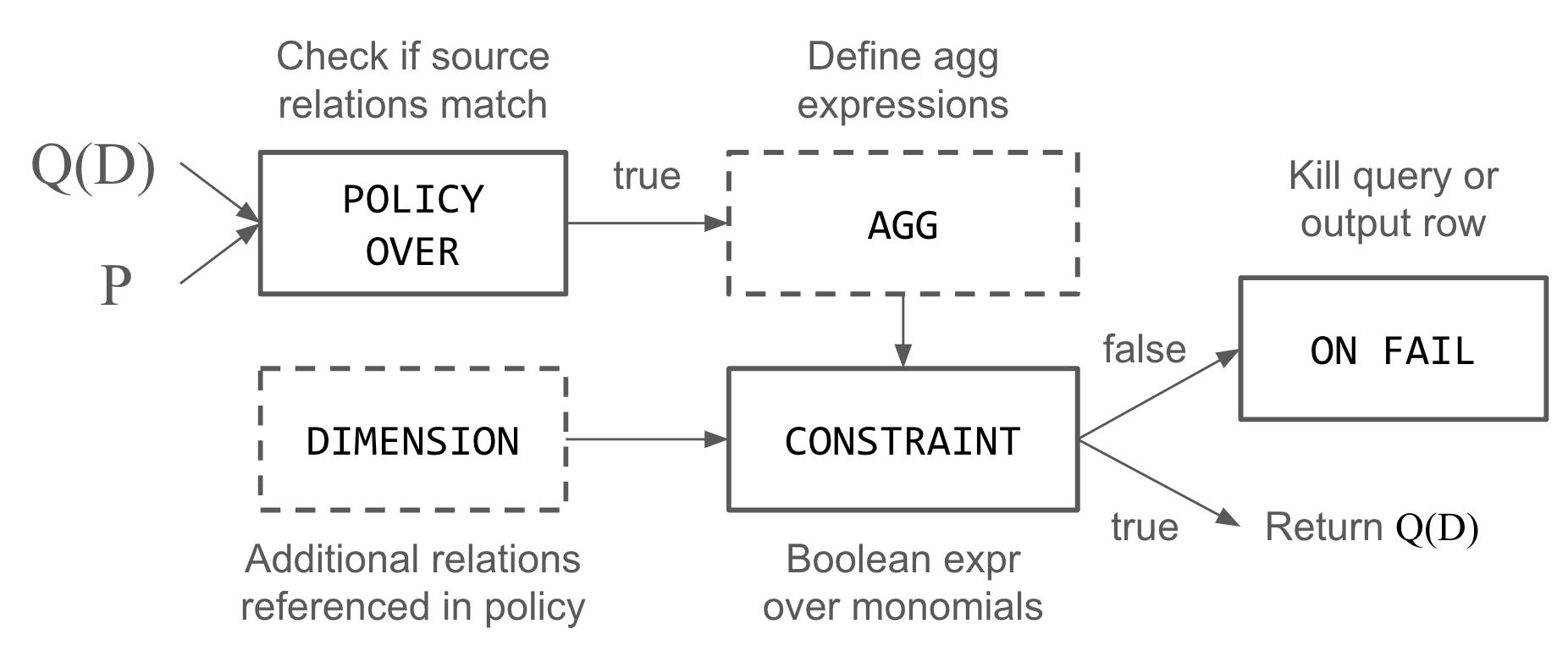}
    \caption{Conceptual evaluation of a \sys policy.  Each box is a clause, labeled arrows are traversed based on the preceding clause's predicate, and dashed boxes are optional.}
    \label{fig:conceptualeval}
\end{figure}

\Cref{fig:conceptualeval} summarizes a \sys policy's components as boxes in its conceptual evaluation.  For brevity, we introduce each clause through the below policy examples.  The business process example also introduces a policy variation to accommodate simple update statements such as those in \Cref{sss:processes}.

\subsubsection{Disaggregation}\label{sss:disaggregationpolicy}
The \code{POLICY} for \Cref{sss:disaggregation} requires that statistics \code{OVER} the \code{students} relation must not aggregate students from a protected class (e.g., between ethnicities).  The \code{POLICY OVER} clause only applies the policy to queries that reference the listed relations (e.g., \code{students}).  The (optional) \code{AGG} clause defines a set of aggregation expressions---\code{count distinct} here---and the \code{CONSTRAINT} clause checks that the aggregation is equal to 1. If not for any output row, the query is \code{KILL}ed by raising an exception in a UDF.

{\small
\begin{verbatim}
  POLICY OVER students
  AGG count(distinct ethnicity) as cnt
  CONSTRAINT cnt = 1
  ON FAIL KILL QUERY
\end{verbatim}
}

\subsubsection{K-anonymity}
This policy ensures that output rows with fewer than 3 unique constituents, and whose query is issued by a public user, should not be released.  The \code{DIMENSION} clause allows the policy to reference additional relations or subqueries, for instance the catalog's \code{users} table.   Unlike \code{POLICY OVER}, \code{DIMENSION} is not used to check whether the policy should apply to a given query.  Here we lookup the current user by id and check that their role is not \code{Public}.  Finally, \code{KILL ROW} ensures the query still returns results, but violating output rows are dropped.


{\small
\begin{verbatim}
  POLICY OVER constituents
  DIMENSION users
  CONSTRAINT users.id = current_user and 
    not (user.role = 'Public' and count(distinct id) < 3)
  ON FAIL KILL ROW \end{verbatim}
}




\subsubsection{Business Processes}
The transitions described in \Cref{sss:processes} are modeled as updates to a relation's attribute.   To manage data flows through updates, we extend the language with an alternative \code{POLICY UPDATE users} clause to denote the updated \code{users} table.  This composes with the existing clauses because the set of updated tuples can be modeled as a subquery that computes the updated attribute expressions.  We then apply the conceptual evaluation in \Cref{fig:conceptualeval} to the subquery.  

For example, the policy below uses \code{DIMENSION} to include the previous \code{users} table.  It ensures that all contributing rows from the previous users table to the updated users table have status `Created', and that the updated status is `Verified'. If not, the violating tuples are not updated.

{\small
\begin{verbatim}
  POLICY UPDATE users as newu
  DIMENSION users as oldu
  AGG bool_and(oldu.status = 'Created') as allc
  CONSTRAINT newu.status = 'Verified' and allc
  ON FAIL KILL ROW
\end{verbatim}
}
\noindent In general, \sys can enforce acyclic FSMs whose state is identified by a single attribute (e.g., \code{users.status}) by translating each transition as described above.

\subsubsection{Prompt Injection}

We now describe a conservative and simple policy to prevent prompt injection. 
Consider a setting where every relation contains a \code{sanitized} attribute.   User- and agent-generated writes are marked as \code{false} by default, and set to \code{true} if a sanitization procedure is applied to the tuple (e.g., user validation, regex check, etc).
This can be enforced by defining a simple policy for every table \code{T} in the database that removes output rows whose input monomials are not all sanitized. 
{\small
\begin{verbatim}
  POLICY OVER T 
  CONSTRAINT bool_and(sanitized) 
  ON FAIL KILL ROW
\end{verbatim} }


%

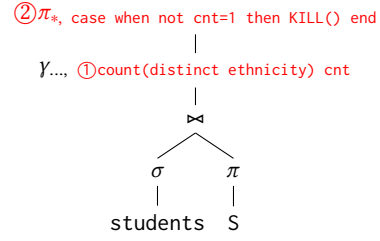
\begin{figure}
    \begin{center}
    \begin{tikzpicture}[level distance=0.7cm]
    \Tree
      [.{\textcolor{red}{\textcircled{2}$\pi_{*,\code{ case when not cnt=1 then KILL() end}}$}}
        [.{$\gamma_{...,\textcolor{red}{\code{ \textcircled{1}count(distinct ethnicity) cnt}}}$}
          [.{$\bowtie$}
            [.{$\sigma$}
              [.{\code{students}} ]
            ]
            [.{$\pi$}
              [.{\code{S}} ]
            ]
          ]
        ]
      ]
    \end{tikzpicture}
    \end{center}
    \caption{The query plan for $\gamma(\sigma(\code{students})\bowtie\pi(S))$ is rewritten to enforce the disaggregation policy in \Cref{sss:disaggregationpolicy}.}
    \label{fig:disaggpolicy}
\end{figure}

\subsection{DFC Policy Evaluation} \label{sec:policy-eval}

\sys policies are enforced by rewriting the user's query $Q$. A naive approach is to use logical provenance capture methods~\cite{glavic2009perm,arab2018gprom}
to rewrite $Q$ to produce provenance annotations as a view, and then append the policy as a post-processing query on top of the view.  However, our goal is {\it not} to capture provenance, which can slow query execution by orders of magnitude, but to efficiently evaluate provenance-based policies which may {\it not} require capturing provenance explicitly.  
We illustrate the key ideas by applying the disaggregation and k-anonymity policies to a join-aggregation query.   We omit query expressions because they do not do not affect the data flow structures.

\subsubsection{Disaggregation}

\Cref{fig:disaggpolicy} illustrates the rewritten query plan for the query $\mathcal{Q}=\gamma(\sigma(students)\bowtie \pi(S))$, with new expressions and operators in \red{red}. In contrast to provenance capture systems like GProM~\cite{glavic2009perm,arab2018gprom},
which instrument every operator to compute and/or propagate provenance annotations, \sys does not modify any of the operators before $\gamma$ either because the operator does not change the output tuple's provenance polynomial (e.g., $\sigma$) or the monomial contribution does not change (e.g., $\bowtie$).   The only modifications is to (1) inline the \code{AGG} clause in the $\gamma$ operator, and (2) a final projection to kill the query if there is a violation.  


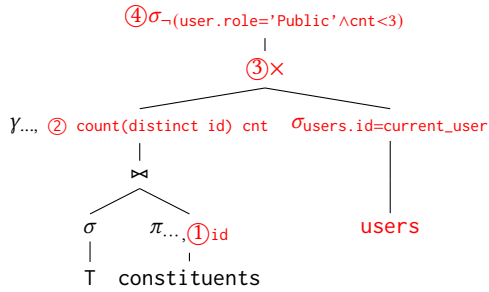
\begin{figure}
    \begin{center}
    \begin{tikzpicture}[level distance=0.7cm]
    \Tree
      [.{\textcolor{red}{\textcircled{4}$\sigma_{\neg(\code{user.role}=\code{'Public'} \land \code{cnt} < 3)}$}}
        [.{\textcolor{red}{\textcircled{3}$\times$}}
          [.{$\gamma_{...,\textcolor{red}{\code{ \textcircled{2} count(distinct id) cnt}}}$}
            [.{$\bowtie$}
              [.{$\sigma$}
                [.{\code{T}} ]
              ]
              [.{$\pi_{\cdots,\red{\textcircled{1} \code{id}}}$}
                [.{\code{constituents}} ]
              ]
            ]
            ]
          [.{\red{$\sigma_{\code{users.id}=\code{current\_user}}$}}
            [{\textcolor{red}{\code{users}}} ]
          ]
        ]
      ]
    \end{tikzpicture}
    \end{center}
    \caption{K-anonymity policy on $\gamma(\sigma(T)\bowtie\pi(\code{constituents}))$.}
    \label{fig:kpolicy}
\end{figure}

\subsubsection{K-anonymity}
\Cref{fig:kpolicy} shows the rewrite for
$\mathcal{Q}=\gamma(\sigma(T)\bowtie\pi(\code{constituents}))$:
(1) the projection includes \code{constituents.id} to later evaluate the policy, (2) the number of distinct constituents is counted, and (3) the aggregation result is augmented with the current user tuple.  The root operator (4) removes output tuples where the current user has a public role and the count is too low.  

\subsubsection{Early Experimental Results}

We evaluate \sys against baselines that capture the query's provenance and then filter the results using the annotations. {\it Logical} uses GProM~\cite{arab2018gprom} to rewrite and optimize the resulting query, while {\it Physical} uses the recent SmokedDuck~\cite{mohammed2023sd} system that modifies DuckDB to capture provenance with near-zero overhead, then calculates counts with FaDE~\cite{mohammed2024fade}.

We consider a subset of TPC-H queries at scale factor 1 that are monotonic and scan the \code{lineitem} table. We add a single policy that drops rows if there are fewer than 1500 total contributors from the \code{lineitem} table. This simple policy is selected to enable FaDE in the {\it Physical} baseline which doesn't support complex aggregations nor predicates.

\begin{figure}
    \centering
    \includegraphics[width=\linewidth]{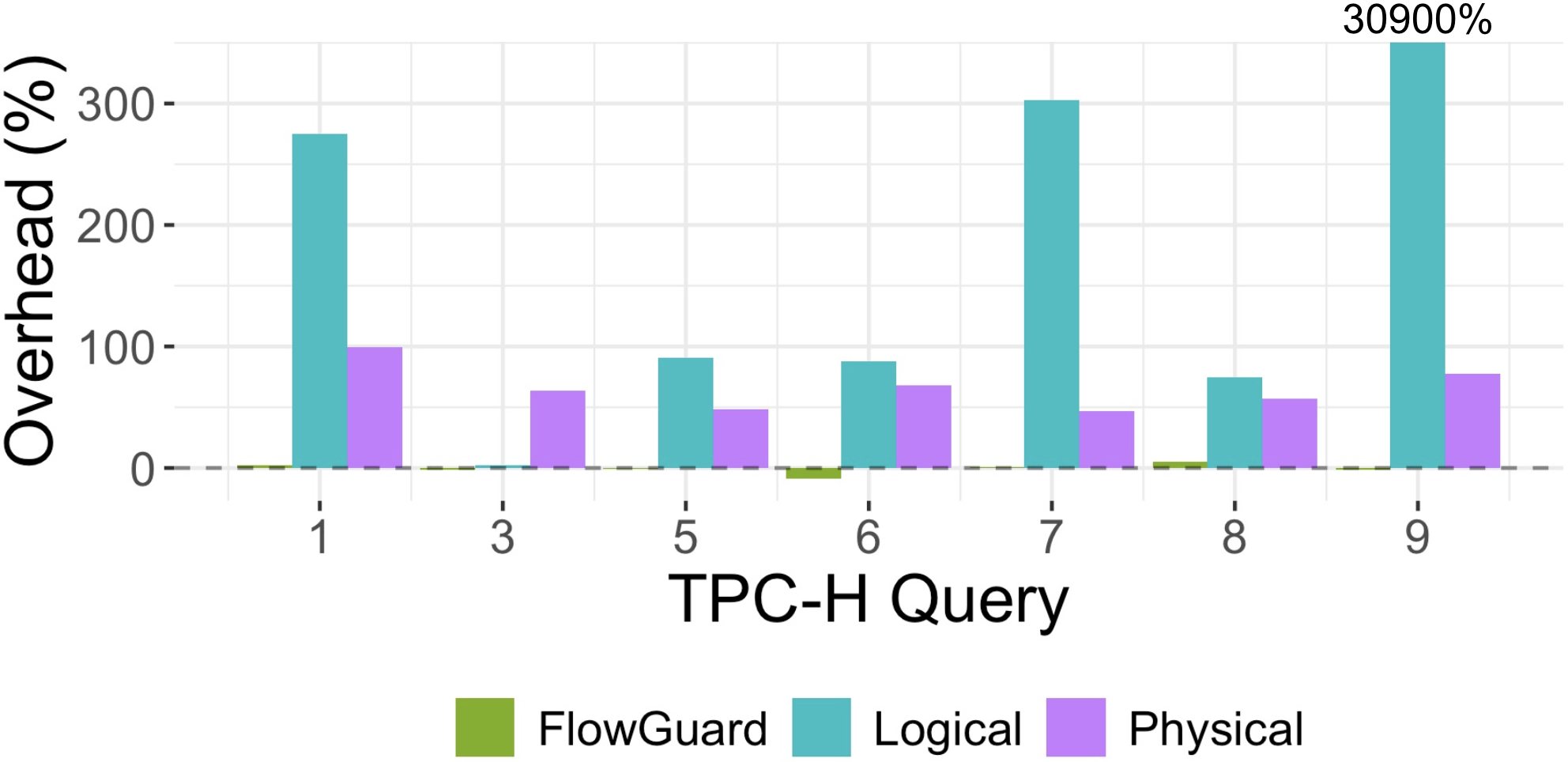}
    \caption{\sys incurs ${\sim}0$ overhead for most queries and is sometimes faster than evaluating the query without a policy.}
    \label{fig:tpch}
\end{figure}
\Cref{fig:tpch} reports the time to execute each TPC-H query and enforce the policy.  Despite the lack of serious optimizations for \sys, we find that \sys consistently reduces policy overhead by one or more orders of magnitude than the alternatives because it completely avoids provenance capture. \sys also does not require modifying system internals, unlike {\it Physical}.

\section{General Data Flow Control}
\label{sec:future}





The policies described until this point have focused on logical data flows within the DBMS. This section outlines the research challenges toward more general data flow control within the DBMS and beyond: potentially supporting agent ecosystems that interact with a multitude of tools.


\subsection{In-DBMS Data Flow Control}
We discuss methods to expand the logical rewrite approach and to extend beyond basic query rewriting.   


\subsubsection{Logical Rewrite Approaches}
\ 

\stitle{Richer Interventions:}
This paper focused on interventions that kill the query altogether or remove an output tuple.   However, a useful variation is to remove an offending input tuple or monomial from the data flow {\it during execution} based on a policy violation.  For instance, conditional tuple access controls such as: ``{\it include all users if \code{ >10} contributors otherwise remove users with \code{opt\_out=true}}''.

A variation of this has been studied in the context of accelerating deletion propagation using provenance polynomials~\cite{mohammed2024fade}, 
but executes post-hoc to the query and does not support non-monotonic blocks that may produce {\it new} intermediate tuples (and thus new data flows) not seen during original query execution.  


\stitle{Scale and Enforcement Overhead:} 
There is potential for administrators to define hundreds or thousands of data flow policies, similar to access control policies~\cite{microsoft2025centralaccess}.   Further, if individual users are empowered to establish their own policies on how their data may be used and combined for downstream applications, then the scale of policies can exceed millions---at the extreme, every user, employee, team, and organization might establish policies.   

Major challenges include how policy enforcement can be administered, how the policy language can be safely composed, how interactions between new and existing sets of policies can be evaluated, and how a large set of policies can be efficiently enforced. For the latter challenge, our early experiments showed that \sys incurs marginal runtime overhead through naive query rewrites that are unoptimized. In principle, the \code{KILL} interventions serve to eliminate data flows during execution, and should be expected to improve query performance if the cost of constraint evaluation is low and the expensive data flows can be filtered early on.

\subsubsection{Beyond Query Rewriting}
We now discuss supporting policies that go beyond the capabilities of basic query rewriting.  






\stitle{Multi-table Policies}: These policies introduce performance challenges that require special attention.  Consider a policy like: \code{POLICY OVER T, S CONSTRAINT median(T.x + S.y) > 10...}
This is difficult to efficiently enforce for queries like $\gamma(\code{T}) \bowtie \code{S}$, which aggregate \code{T} before \code{T.x + S.y} can be calculated.   A naive approach logically rewrites the query to capture provenance annotations for the subplan  containing $T$ and $S$, and then evaluates the constraint post-join.  However, based on our early experiments, this degrades performance considerably.   A promising approach is to limit multi-table policies to constraints over e.g., semiring aggregates such as \code{SUM} or \code{COUNT} that can be distributed through the join.

\stitle{Multi-query and Multi-step Enforcement:}  
In practice, agents and data analysts work iteratively to read and write data. While no individual query may trigger a policy violation, the sequence of queries and their dependencies form a large data flow that could violate a policy. This is a common source of prompt injection (\Cref{sec:prompt-injection}),  where unsanitized data circuitously propagates through intermediate queries that is eventually added to a LLM prompt.  

The core challenge is, given a DAG of already-executed queries, when is it possible to infer violations or non-violations for any later queries that may be appended to the DAG?   A promising direction is to focus on monotonic constraints, for which a violation of a threshold constraint like \code{min(x)>10} is guaranteed to persist independent of future queries. 
This can benefit from recent work on leveraging lattice structures in distributed data programs~\cite{power2025free}.


\stitle{Physical Interventions:}
The interventions discussed so far can be modeled using killing UDFs and filter operators.  However, there are many cases where it is helpful to physically intervene on the data flow, perhaps to log, suppress, or even change tuples mid-flight. 
For instance, it can be useful to log data flows that are ``near'' policy violation boundaries for double checking, to route or buffer violating tuples for human verification during execution, to cap intermediate result sizes to avoid unexpected resource consumption from poorly written queries, or to merge groups that are too small to obey K-anonymity requirements.   

\begin{figure}
    \centering
    \includegraphics[width=\linewidth]{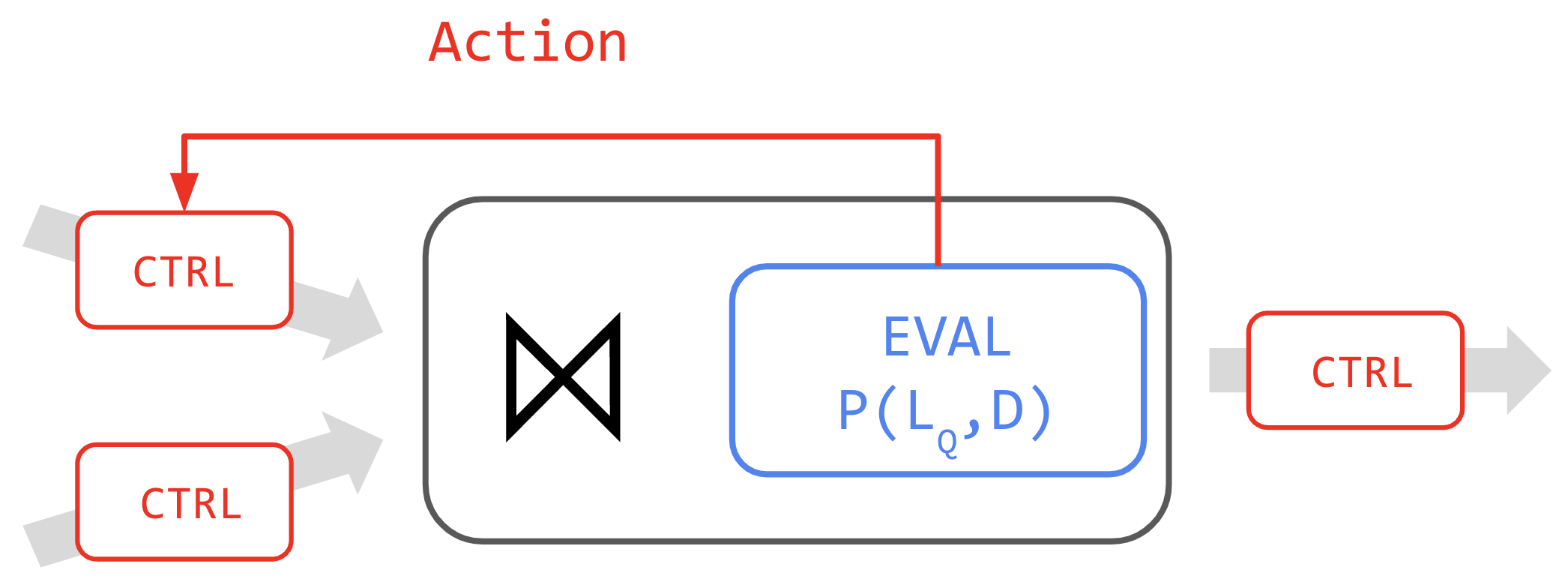}
    \caption{Lightweight \red{\code{CTRL}} and \blue{\code{EVAL}} operators check and intervene on data flows for a policy $\blue{P(L_\mathcal{Q}, \mathcal{D})}$ over the query $\mathcal{Q}$'s data flow $L_\mathcal{Q}$ and database $\mathcal{D}$, and intervention \red{\code{Action}}.}
    \label{fig:ctrl}
\end{figure}

A promising approach (\Cref{fig:ctrl}) is to introduce a pair of light-weight \red{\code{CTRL}} and \blue{\code{EVAL}} operators.  \blue{\code{EVAL}} checks policy constraints and sends signals to \red{\code{CTRL}} operators, which sit between relational operators, to redirect or change data flow. The major challenge is the logical-physical tension described in \Cref{sec:design-challenge} and designing the API to expose {\it physical} interventions to a logical policy language.

\subsection{From DBMS Tools to General Tools}

While the preceding sections focused on an in-DBMS solution, agentic systems interact with an ecosystem of tools and other agents. We believe that modeling the tools from a relational perspective can help apply these ideas to new settings.  However, it raises questions such as: What DFC capabilities do tools need to expose? How can global policies be expressed? And how to operate gracefully when tools do not support DFC or support data flow tracking imperfectly?

Data flow itself can be viewed as a form of observability for data-centric programs. The core challenge in generalizing beyond the DBMS is that existing observability platforms follow different semantics: the DBMS is data-centric; OS systems are process-centric, concerned with inter-process communication; the network is packet-centric, focused on network flows; modern observability tools are service-centric, focused on requests; and agent frameworks are token-centric.  These semantics do not readily translate between one another, and the fragmentation makes it difficult to establish a coherent model for end-to-end data flows.   The central challenge, therefore, is developing a mechanism to bridge these conceptual divides and create a unified, cross-domain perspective on data flow that is both efficient and understandable.

Another challenge is designing a tool interface for data flow control in the context of Model Context Protocol. Exposing full fine-grained provenance is unrealistically expensive and can overwhelm agent contexts. One possibility is that tools declare a schema for the underlying data flow to support discovery and registration, and expose subsets of a common DFC policy language to be enforced. Access to tools can take the degree of DFC visibility into account.  This raises theoretical and language design challenges in creating a policy language friendly to federated declaration and enforcement.


\section{Conclusion} 

We advocated for Data Flow Control (DFC) to safely deploy LLM agents in stateful environments. Shifting enforcement of data flow policies to underlying data systems decouples administrative policies from agent development, and automatically mitigates risks like regulatory violations, process corruption, and prompt injection. We presented an early instantiation of a rewrite-based policy language called \sys that incurs low overhead. We then outlined research challenges towards DFC for general agent ecosystems. We believe that enhanced systems capabilities will let the vibes flow.

\begin{acks}
This work was supported by the National Science Foundation (1845638 (CAREER), 1740305, 2008295, 2106197, 2103794, 2312991), DAPLab partners (Intellect Design, Amazon, Tidalwave, Veris, Infosys) and gifts from Amazon, Google, Adobe, CAIT.
\end{acks}

\bibliographystyle{ACM-Reference-Format}
\bibliography{main}

\end{document}